\documentclass[10pt,a4paper]{article}
\usepackage{etex}
\usepackage[utf8]{inputenc}
\usepackage[TS1,T1]{fontenc}
\usepackage{xspace}
\usepackage[a4paper,tmargin=3truecm,bmargin=3truecm,rmargin=2.2truecm,lmargin=2.2truecm]{geometry}
\usepackage{amssymb,amsmath,mathtools}
\usepackage[all]{xy}
\usepackage{enumerate}
\usepackage[english]{babel}
\usepackage{hyperref}
\usepackage{lmodern}
\usepackage{array,booktabs}

\usepackage[shadow,backgroundcolor=red!20]{todonotes}

\newcounter{mnotecount}[section]

\usepackage{amsthm,thmtools}

\newtheorem*{thm}{Theorem}
\newtheorem*{lem}{Lemma}
\newtheorem*{cor}{Corollary}

\theoremstyle{definition}

\theoremstyle{remark}

\hypersetup{
plainpages=false,
colorlinks=true,
linkcolor=black, 
anchorcolor=black, 
citecolor=red, 
urlcolor=black, 
menucolor=black, 
filecolor=black, 
bookmarksopen=true,
bookmarksnumbered=true}

\newcommand\gR{{\mathbb R}}

\DeclareMathOperator{\Ad}{Ad}

\hypersetup{
pdfauthor={J. Attard, S. Lazzarini},
pdftitle={Weyl Invariance in gravity and the Wess-Zumino functional},
pdfsubject={},
pdfcreator={pdflatex},
pdfproducer={pdflatex},
pdfkeywords={}
}

\begin{document}


\begin{titlepage}

\title{A Note on Weyl invariance in gravity and the Wess-Zumino functional \footnote{Dedicated to the memory of Raymond Stora (1930-2015).}\\[1cm]}

\author{J. Attard and S. Lazzarini \footnote{Corresponding author: {\tt serge.lazzarini@cpt.univ-mrs.fr}.}}
\date{}

\maketitle

\vskip -3mm
\centerline{Aix Marseille Univ, Univ Toulon, CNRS, CPT, Marseille, France.}



\vskip 2cm

\begin{abstract}
It is shown that the explicit calculation of the Wess-Zumino
functional pertaining to the breaking term of the Weyl symmetry for
the Einstein-Hilbert action allows to restore the Weyl symmetry by
introducing the extra dilaton field as Goldstone field. 
Adding the Wess-Zumino counter-term to the Einstein-Hilbert action reproduces the usual Weyl invariant action used in standard literature. Further consideration might confer to the Einstein-Hilbert action a new status.
\end{abstract}




%

\end{titlepage}





\section{\label{sec-introduction} Introduction and motivation}

It is fair to say that the conformal Weyl symmetry which induces a local rescaling of a metric $g_{\mu\nu}(x) \mapsto \Omega^2(x) g_{\mu\nu}(x)$ is still a fascinating local symmetry. As such, it can be considered on the same footing as a gauge symmetry, a standpoint we shall adopt in the paper.

\smallskip
Quite recently G. `t Hooft in \cite{Hooft:2015rdz} 
put an emphasis on the use of what he called the local conformal symmetry, and the rôle of a dilaton field with only renormalizable interactions, in particular when gravity couples to matter. Some years before, in \cite{Gover:2008pt} Weyl symmetry is shown to be related to the origin of mass. As `t~Hooft says, local conformal symmetry is not well understood yet. Let us add in particular, its relationship with scale (or dilation) symmetry $x^\mu \mapsto \lambda x^\mu$ which confers the canonical dimension to a field. This canonical dimension can be reinterpreted as a weight when fields are geometrically considered as densities. 

\smallskip
Both \cite{Gover:2008pt,Hooft:2015rdz} used the conformal Weyl invariant local functional in gravitation in $n$-dimensional spacetime 
\begin{align}
S(g,\sigma) = -\frac{1}{2}\int \sqrt{|\det g|}\, \sigma^{-n}\Big(\,(n-2)(n-1)||\nabla\sigma||^2 + R(g) \, \sigma^2\Big) d^n x
\label{S-W-inv}
\end{align}
where $||\nabla\sigma||^2 = g^{\mu\nu} \partial_\mu \sigma \partial_\nu \sigma$ is the Riemannian scalar product.
The Weyl invariance is achieved when beside the Weyl rescaling of the metric $g$, the scalar field $\sigma$ transforms according to $\sigma \mapsto \Omega \sigma$. The latter is a Weyl compensating field (or dilaton) in order to restore Weyl invariance of the theory while maintaining the locality principle for the functional in the fields $g$ and $\sigma$.

\smallskip
Former approaches showed that non-Weyl invariant theories (as the Einstein-Hilbert action for gravity is) can be turned into Weyl invariant ones by using a Weyl compensating scalar field.
According to the existing literature on the subject, it would appear that this functional action was considered at the beginning of the 1970's by B. Zumino \cite[in particular, reference 13 therein]{Zumino:1970tu} and S. Deser \cite{Deser:1970hs}, independently. This was around the same period of the occurrence of the so-called Wess-Zumino term \cite{WZ71}. Surprisingly enough, no relationship with the latter was put into evidence 
at that time and subsequently \cite[formula (14)]{Wess73}, according to the best of our knowledge. 

\smallskip
Relying onto the standpoint of locality principle in QFT and considering the metric $g$ as an external gravitational field,
we shall adopt the attitude to consider the breaking of the Weyl rescalings (conformal transformations) of the Einstein-Hilbert action as an ``anomaly''. If it turns out to be the case, by making use of the BRST techniques, namely, it is a BRST 1-cocyle from which the Wess-Zumino term can be constructed. Recall that the latter integrates the anomalous term and can be used as counter-term to restore the symmetry. The price to pay usually is to add in the theory a new Goldstone field which carries a non linear transformation law.

\smallskip
This legitimately raises the natural question whether the introduction of the above dilaton field $\sigma$ as compensating field pertaining to the Weyl conformal symmetry, and accordingly, the modification of the Einstein-Hilbert action into the action \eqref{S-W-inv}, stems from the usual generic construction of a Wess-Zumino term (or action or functional). 
This issue will be addressed in the present paper whose purpose is to provide a positive answer to this somewhat conceptual issue. 

\smallskip
The construction of the Wess-Zumino functional as in \cite{WZ71}, \cite{Sto84}, 
\cite[formula~(4.33)]{Zumi84a}, \cite{Zum85} and \cite[see p.~164]{ManeStorZumi85a} is well-known for usual gauge theories. That is, when the gauge anomaly is of polynomial type, {\em i.e.} constructed through the so-called descent equations stemming from an invariant polynomial given by a characteristic class of the underlying principle fiber bundle. What about the construction of a Wess-Zumino functional in the case where the anomaly is not ``polynomial'', namely, there is no $Q$ coming from the Chern-Simons transgression formula?
For instance, diffeomorphism anomalies, Weyl anomaly fall into this category.
The generic construction given in \cite{Sto85,
Sto93} takes fully into account both the cases.
Let us call that construction the ``Stora construction''\footnote{Historically and according to the best of our knowledge, a preliminary proof of the Stora construction was given as an appendix in~\cite[appendix F]{Laz90} along the Stora's ideas and used for integrating the 2-D Weyl anomaly in \cite{KLT90}. The proof was completed in \cite{Sto93} by Stora himself. 
}. We shall apply it in the context of local conformal symmetry in gravity. In order to achieve this goal, the general BRST treatment of anomalies will be used. 

\smallskip
The paper is organised as follows. Section \ref{sec:Weyl-sym} will serve to fix the notation.
In section~\ref{sec:Weyl-breaking}, we shall show that the breaking term inferred by the Einstein-Hilbert action falls into the usual algebraic BRST cohomology.
In section~\ref{sec:WZ-term} the corresponding Wess-Zumino term will be computed explicitly.
The paper is closed by some concluding remarks and open questions.
Two appendices are devoted to the construction of the Wess-Zumino action according to Raymond Stora.


\section{\label{sec:Weyl-sym} Weyl rescalings}

Let $M$ be a $n$-dimensional smooth manifold and consider ${\cal M}et(M)$ the infinite dimensional space of (pseudo-)Riemannian metrics on $M$.  

A Weyl rescaling is a mapping on ${\cal M}et(M)$ defined by $g(x) \mapsto \varphi(x) g(x)$ where $\varphi\in C^\infty(M,\mathbb{R}^*_+)$ is a smooth positive function defined on $M$. It is also named a conformal change of the metric $g$. Such transformations yield an abelian transformation group acting on the space ${\cal M}et(M)$. 

We shall parametrize for convenience $\varphi(x) = \Omega^2(x) = e^{2\phi(x)}$ so that the Weyl rescaling of a metric reads
\begin{align}
\label{W-of-g}
& g\rightarrow \bar{g} = \Omega^2 g \, , &  \delta_\text{\tiny Weyl} g =: \delta_\phi\, g = 2\phi\, g &
\end{align}for respectively, a finite local conformal change of the metric $g$, or infinitesimally linearized version. Notice that $[\delta_\phi,\delta_{\phi'} ] = \delta_{[\phi,\phi']} \equiv 0$ due to the abelian feature of the Lie algebra $(\mathbb{R},+)$ of the Weyl gauge group.

\smallskip
Accordingly, the scalar curvature $R(g)$ transforms under a conformal rescaling as (see {\em e.g.} \cite[§ 3.9]{Gold62})
\begin{align}
\label{W-of-R}
R(\bar{g}) = e^{-2\phi} \left( R(g) + 2(n-1) \Delta \phi - (n-1)(n-2) ||\nabla\phi||^2 \right)
\end{align}
or under an infinitesimal Weyl transformation as
\[
\delta_\text{\tiny Weyl} R(g) = 2(n-1) \Delta\phi - 2\phi R(g)
\]
where the Laplacian acting on scalar functions is given by 
(see {\em e.g.} \cite[§ 2.7]{Gold62})
\[
\Delta = - \nabla_\mu \nabla^\mu = \frac{-1}{\sqrt{|\det g|}}\, \partial_\mu \left( g^{\mu\nu} \sqrt{|\det g|}\ \partial_\nu \right).
\]

\smallskip
If one considers the quotient space ${\cal M}et(M)/C^\infty(M,\mathbb{R}^*_+)$, this is the infinite dimensional space of the conformal classes $[g]$ of the metrics. Morally, a Weyl (scale) invariant physical theory is supposed to depend only on the conformal classes of metrics. Considering the local Weyl rescalings as a gauge group, one can pass to the quotient space
${\cal M}et(M) \stackrel{\text{Weyl}}{\longrightarrow} {\cal M}et(M)/C^\infty(M,\mathbb{R}^*_+)$.
However, as Raymond Stora used to insist on, owing to the locality principle, the (gauge) Weyl invariant action functional must depend on a representative of the conformal class, with the inherent problem of the ambiguity on the choice of this representative, say $g$. This amounts one to working on the space $\Gamma_\text{\tiny loc}({\cal M}et(M))$ of local functionals in the metric. However, in gauge theory, observables are functions on orbit space, or equivalently, gauge invariant functions on field space (which is here ${\cal M}et(M)$) \cite{Stora:1995er}.

\smallskip
For the local conformal symmetry the corresponding (classical) Ward identity acting on local functionals reads
\begin{align}
\label{WI}
\int_M d^n x\, \delta_\text{\tiny Weyl} g_{\mu\nu}(x) \frac{\delta \Gamma[g]}{\delta g_{\mu\nu}(x)}   = \int_M d^n x\, 2\phi(x) g_{\mu\nu}(x) \frac{\delta \Gamma[g]}{\delta g_{\mu\nu}(x)} = 0.
\end{align}

\smallskip
Let us now consider the Einstein-Hilbert action which is such a local functional of $g$
\begin{align}
S_\text{\tiny EH}(g) = \int_M \text{vol}(g) \, R(g) = \int_M d^n x \sqrt{|\det g|}\, R(g)
\end{align}
where $\text{vol}(g)$ is the volume form on $M$ and $R(g)$ is the scalar curvature, both associated to the metric $g$.

\smallskip
The next step will be the study of the behavior of the Einstein-Hilbert action under local conformal rescalings.

\section{\label{sec:Weyl-breaking} Another Weyl ``anomaly''}

As is well known, the Einstein-Hilbert gravitation theory yields a spontaneous breaking of the conformal Weyl symmetry. We shall study this breaking of conformal symmetry in the framework of the BRST differential algebra \cite{BRS76,Sto77}. Thus, turning the Weyl parameter $\phi$ into the Faddeev-Popov ghost field, still denoted by $\phi$ with $\phi^2=0$ (abelian Lie algebra), the corresponding Slavnov operation $s$ acting on the field generators --see {\em e.g.} \cite{Bandelloni:1981zp,Baulieu:1986hw}-- is defined by
\begin{align}
\label{W-BRS}
s g &= 2\phi g, & s \phi &= 0, & \text{with } s^2 &=0 .
\end{align}
In particular, one has
\begin{align}
\label{W-var}
s \sqrt{|\det g|} &= n \phi \sqrt{|\det g|}, & s R(g) &= 2(n-1) \Delta\phi - 2\phi R(g).
\end{align}
This Slavnov operation $s$ will act on the functional space $\Gamma_\text{\tiny loc}({\cal M}et(M))$.
A direct computation shows that the variation of the Einstein-Hilbert action under Weyl transformations is infinitesimally given by
\begin{align}
s S_\text{\tiny EH}(g) = A(\phi,g)
\end{align}
with
\begin{align}
\label{anom}
A(\phi,g) = \int_M  
d^n x \sqrt{|\det g|}\, \Big( (n-2) \phi R(g) + 2(n-1) \Delta \phi \Big).
\end{align}
This means that the Ward identity \eqref{WI} is already broken at the classical level by the Einstein-Hilbert action.
By inspection, the functional $A(\phi,g)$ is local in $g$ and linear in the ghost argument $\phi$, and it turns out to fulfill the celebrated Wess-Zumino consistency condition~\cite{WZ71,Sto77} in its BRST formulation, namely,
\begin{align}
\label{WZconsistency}
s A(\phi,g) = 0
\end{align}
as it can be checked explicitly. In the course of the computation, an integration by parts must be performed in order to get an integrand of the type $||\nabla\phi||^2$ which vanishes by Faddeev-Popov argument.

\smallskip
Hence, since \eqref{WZconsistency} expresses a 1-cocycle condition, one can analogously treat $A(\phi,g)$ as an ``anomaly'' even if it is not a quantum breaking of the Weyl symmetry. Here, it is rather a geometrical breaking. However, one might speculate that the Einstein-Hilbert action could be considered as a local ``vacuum functional'' depending on an external gravitational field $g$. The latter ought to come from a field theory coupled to a gravitational field, similarly to the approach followed in~\cite{Baulieu:1986hw,Bec88} for the bosonic string in the 2D case, where by a gauge fixing condition the gravitational field becomes external. Let us also mention that the Einstein-Hilbert action may be seen as resulting from the spectral action principle initiated in \cite{Chamseddine:1996zu}
together with the Standard Model Lagrangian (at least at the tree level).

\smallskip
The main issue now is to restore the local Weyl conformal symmetry by proceeding along the line used in gauge theory to reabsorb the anomaly.

\section{\label{sec:WZ-term} The Wess-Zumino functional}

The reader is recalled that the Wess-Zumino functional is defined to integrate the anomaly, see {\em e.g.} \cite{ManeStorZumi85a,Sto93} for a BRST treatment of this issue, that is (according to our situation at hand) 
\begin{align}
\label{sGam}
s \Gamma_\text{\tiny WZ} = - A(\phi,g).
\end{align}
In fact, the anomaly is trivialized at the cost of introducing an extra field with values in the Lie algebra of the gauge group.
Two appendices give a detailed account on Raymond Stora's ideas on the construction of the Wess-Zumino action as an important item in gauge theory.\footnote{The minus sign in formula \eqref{sGam} is introduced for convenience, contrary to the general construction given in \ref{WZS}.}

\medskip
In our case, the anomaly does not seem to be a ``polynomial anomaly'' (namely of Adler-Bardeen type). A priori, there is no Chern-Simons transgression formula. However, as said in section
\ref{sec-introduction}, there is still a way to construct the Wess-Zumino functional~\cite{Sto93}. 

\smallskip
To this end, take a 1-parameter subgroup for $t\in[0,1]$, $ \gamma_t = e^{-t\tau}$ in the Weyl gauge group from the identity element $ \gamma_0=1$ (constant function) to the positive function $ \gamma_1 = e^{-\tau} =: \gamma $. Accordingly, its action on a given metric $g$ defines a path in the conformal class $[g]$ of the metric $g$ (namely, in the Weyl gauge orbit of the latter)
\begin{align}
\label{homotop-sur-g}
g_t &=  \gamma_t^2 g = 
e^{-2t\tau} g \, , \qquad \text{with }  g_0 = g \mbox{ and } g_1 = \gamma^2 g = e^{-2\tau} g.
\end{align}
Consider the pull-back to the interpolating family $\gamma_t$ of the Maurer-Cartan form on the gauge group associated to the Weyl group
\begin{align}
\label{MC}
\omega_t =  \gamma_t^{-1} d_t  \gamma_t = -2\tau dt
\end{align}
which is found to be no longer depending on the $t$-parameter.
Following \cite{ManeStorZumi85a,Sto85} one extends the Slavnov operation $s$ to the added scalar field $\tau$ by requiring
\begin{align}
\label{sg1}
s g_1 = s (e^{-2\tau} g) = 0 \Rightarrow s \tau = \phi.
\end{align}
It is like a gauge fixing on the conformal component of the metric field. Rather, it might be viewed as a change of variables within the field space which $g$ and $\tau$ belong to \cite{GaugeInv}. Hence, the field $\tau$ carries 1 as Weyl conformal weight and will play the rôle of dilaton as we shall see.

\medskip
Raymond Stora \cite{Sto93} defines the Wess-Zumino functional, $\Gamma_\text{\tiny WZ}$, by integrating over the interpolating family the anomalous term according to 
\begin{align}
\Gamma_\text{\tiny WZ}(\tau,g) = \int_0^1 A(\omega_t,g_t) = \int_0^1 dt A(-2\tau,g_t)
\end{align}
upon using \eqref{MC}. One has to perform the integration over $t$ of the functional
\[
A(-2\tau, g_t) = - 2 \int_M  
d^n x \sqrt{|\det g_t|}\, \Big( (n-2) \tau R(g_t) + 2(n-1) \Delta_t \tau \Big).
\]
The overall factor $2$ in the r.h.s. comes from the linearity of $A$ in its first argument. Upon using \eqref{W-of-R} and \eqref{homotop-sur-g} and after some algebra, one gets the expression
\begin{align*}
\Gamma_\text{\tiny WZ}(\tau,g) =  -2 \int_M  d^n x \sqrt{|\det g|} 
\int_0^1 dt\ e^{(2-n)t\tau} \Big( &
(n-2)\tau R(g) + 2(n-1) \Delta\tau \\
& + 2(n-2)(n-1)( ||\nabla\tau||^2 - \tau \Delta\tau )t \\
& - 2(n-2)(n-1)^2 ||\nabla\tau||^2 t^2 \Big). 
\end{align*}
Performing the integration over $t$, the corresponding Wess-Zumino functional reads
\begin{align}
\Gamma_\text{\tiny WZ}(\tau,g) =  2\! \int_M d^n x \sqrt{|\det g|} 
\left( e^{(2-n)\tau} \Big( R(g) - 2(n-1) \Delta\tau  
- (n-2)(n-1) ||\nabla\tau||^2 \Big) - R(g) \right).
\end{align}
Upon defining $\sigma=e^\tau\, (= 1/\gamma)$ (the inverse element to $\gamma$) with $s\,\sigma = \phi\sigma$, \footnote{This means, at the finite level $\sigma \rightarrow \bar{\sigma} = \Omega \sigma$.}
as a compensating field of conformal weight one,
the Wess-Zumino action can be rewritten as
\begin{align*}
\Gamma_\text{\tiny WZ}(\tau,g) = 2\! \int_M d^n x \left( \sigma^{2-n} \sqrt{|\det g|} \,
\Big( R(g) - 2(n-1) \Delta\tau  
- (n-2)(n-1) ||\nabla\tau||^2 \Big)
- \sqrt{|\det g|} R(g)  \right)
\end{align*}
and next by using once more the Weyl transformation 
\eqref{W-of-R} for $g \mapsto \sigma^{-2} g$, one is led to
\begin{align*}
\Gamma_\text{\tiny WZ}(\tau,g) = 2\! \int_M d^n x \left( \sqrt{|\det (\sigma^{-2} g|}\, R(\sigma^{-2} g) 
- \sqrt{|\det g|}\, R(g)  \right)
\end{align*}
which is nothing but the difference
\begin{align}
\label{WZ-funct}
\tfrac{1}{2}\, \Gamma_\text{\tiny WZ}(\tau,g) = S_\text{\tiny EH}(\sigma^{-2} g) - S_\text{\tiny EH}(g).
\end{align}
Since the rôle of the counter-term $\Gamma_\text{\tiny WZ}$ is to cancel the anomaly, one thus has by construction
\begin{align}
s\, S_\text{\tiny EH}(\sigma^{-2} g) = s \left( \tfrac{1}{2} \Gamma_\text{\tiny WZ}(\tau,g) + S_\text{\tiny EH}(g) \right) = 0
\end{align}
which is consistent actually with the constraint \eqref{sg1}.
Finally, this yields the Weyl invariant local functional
\begin{align}
\label{S-EH-inv}
S_\text{\tiny EH}(\sigma^{-2} g) = S_\text{\tiny EH}(g) + \tfrac{1}{2} \Gamma_\text{\tiny WZ}(\tau,g) &= \int_M d^n x \sqrt{|\det (\sigma^{-2} g|}\, R(\sigma^{-2} g) \notag\\[-2mm]
\\[-1mm]
& = \int_M d^n x \sqrt{|\det g|} \, \sigma^{2-n} 
\Big( R(g) - 2(n-1) \Delta\tau  
- (n-2)(n-1) ||\nabla\tau||^2 \Big). \notag
\end{align}
Remembering that $\sigma = e^\tau$, one can check that this formula, up to an integration by parts and up to an overall factor $-\tfrac{1}{2}$, is nothing but the Weyl invariant action \eqref{S-W-inv} given in \cite{Zumino:1970tu,Deser:1970hs} and used in \cite{Gover:2008pt,Hooft:2015rdz}.

\smallskip
At this stage some comments are in order.

\section{\label{sec:comments} Comments and outlook}

On the one hand, since $s(\sigma^{-2} g)=0$, it turns out to be obvious that $s\, S_\text{\tiny EH}(\sigma^{-2} g) = 0$.
But, on the other hand, the Weyl invariance has been restored by mimicking a construction coming from Lagrangian gauge theory, by adding the Wess-Zumino counter-term to the well-known Einstein-Hilbert action and thus introducing the so-called compensating field $\sigma$ in order to be consistent with the locality principle.\footnote{Even if this sounds quite simple at first sight, it is grounded on strong QFT principles. Recall that the construction presented in the paper does not appear to have been done in standard literature.} In this respect, and to parallel some Raymond Stora's viewpoints (see section \ref{sec:Weyl-sym}) this rises the following question: Does the Weyl invariant combination $\sigma^{-2} g$ gives a substitute to parametrize the conformal class $[g]$ (of the metric $g$) compatible with the locality principle?
And to ``mirror'' \cite{Hooft:2015rdz}: Does the construction of the Wess-Zumino term provides a canonical way to isolate the dilaton component of the metric?

\medskip
Moreover, according to \cite[p.464]{Zumino:1970tu} an action which is invariant under both Einstein and Weyl symmetries is invariant in the Minkowski flat limit under the 15-dimensional conformal group.
One is led to make contact also with \cite{Bandelloni:1981zp,Baulieu:1986hw} , and one may also remark that if it would be possible to set $\sigma = (\det g)^{1/2n}$ by gauge fixing or as an equation of motion, then $s\, \sigma = \phi \sigma$, since $s\, (\det g) = 2n\phi (\det g)$.
Accordingly, $\sigma^{-2} g = (\det g)^{-1/n} g$ and $\det(\sigma^{-2} g)=1$ so that $\sigma^{-2} g = \hat{g}$, the associated unimodular matrix to the metric $g$, will also serve as a representative of the conformal class $[g]$ of the metric $g$.
It ought to be useful to investigate more in that direction.

\smallskip
Going back to the construction of the Wess-Zumino action, in particular the rôle of the interpolating family, formula \eqref{WZ-funct} can be simply recast as
\begin{align}
\tfrac{1}{2}\, \Gamma_\text{\tiny WZ}(\tau,g) = \int_0^1 d_t S_\text{\tiny EH}(\sigma^{-2t\tau} g) =  \int_0^1 \frac{\partial S_\text{\tiny EH}(\sigma^{-2t\tau} g)}{\partial t} \, dt  = \int_0^1 d_t S_\text{\tiny EH}(g_t) \ .
\end{align}
Not only the integration over the family can be explicitly performed for computing the Wess-Zumino action, but it highlights the Einstein-Hilbert action since one may write 
\begin{align}
\label{Atrivial}
A(\omega_t,g_t)  = 2\, d_t S_\text{\tiny EH}(g_t)
\end{align}
along a path in the gauge orbit given by the conformal class $[g]$ of the metric $g$.
Following Stora's tricks \cite{Sto93}, at the algebraic level (here, the Lie algebra is abelian) the Maurer-Cartan equation $d_t \omega_t =0$, together with $d_t g_t = - d_M \omega_t$ ($d_M$ is the de Rham differential on spacetime $M$), yields
a differential algebra which is similar to the BRS one given in \eqref{W-BRS}. The Wess-Zumino consistency condition leads to $d_t A(\omega_t,g_t) = 0$.
Since $d_t^{\,2}=0$, equation \eqref{Atrivial} indicates that the Wess-Zumino term turns out to be independent of the interpolating family up to smooth deformations and depends only on the bounds. 

\smallskip
Thus $S_\text{\tiny EH}(g_t)$ interpolates along a family of conformally related metrics to the metric $g$. That is within a fiber of ${\cal M}et(M)$ with respect to the conformal symmetry. This might indicate that the Einstein-Hilbert action plays a special role in the construction. If one was able to pass to the quotient space ${\cal M}et(M) / C^\infty(M,\gR^*_+)$, as configuration space, the ``genuine'' physical theory ought to depend only on the conformal classes $[g]$ of the metrics $g$. More investigation deserves to be performed in this matter.

\section*{Acknowledgments}

One of us (SL) has got the great honor of being the last Raymond Stora's PhD student and would like, with this paper, to pay tribute to the memory of Raymond.

The authors apologize 
for references that have been implicitly quoted in the bibliography.

We would like to thank C.~Rovelli and T.~Schücker for fruitful discussions at the earlier stages of this work and for pointing out to us the G.~`t~Hooft's considerations in this matter, in particular reference~\cite{Hooft:2015rdz}. We are indebted to T.~Masson for helpful remarks on the elaboration of \ref{appendix:B}. We also thank T.~Krajewski for drawing our attention to the approach given in~\cite{Iglesias2013}.

\smallskip
This work has been carried out in the framework of the Labex ARCHIMEDE (Grant No. ANR-11-LABX-0033)
and of the A*MIDEX project (Grant No. ANR-11-IDEX-0001-02), funded by the “Investissements d’Avenir”
French Government program managed by the French National Research Agency (ANR).

\appendix
\renewcommand{\thesection}{Appendix \Alph{section}}
\renewcommand{\theequation}{\Alph{section}.\arabic{equation}}

\section{The Stora construction of the Wess-Zumino action \label{WZS}}
\setcounter{equation}{0}

In \cite{Zumi84a,Zum85}, \cite[Section III]{Sto84} or \cite{ManeStorZumi85a,Manes:1984gk,Sto85} the construction of the Wess-Zumino functional was mainly performed in the case where the gauge anomaly was of polynomial type, {\em i.e.} constructed through the so-called descent equations stemming from an invariant polynomial coming from a characteristic class of the underlying principle fiber bundle. What about the construction of a Wess-Zumino functional in the case where the anomaly does not come from a polynomial, namely there is no local functional 
coming from the Chern-Simons transgression formula.
For instance, diffeomorphism anomalies, Weyl anomaly or presently, the ``anomalous'' term obtained from the Einstein-Hilbert action as a spontaneous breaking of the conformal Weyl invariance, they all fall into this latter type. 

In this appendix, we would like to report on a more general construction which deals such a situation, even more, with any generic situation, for any Lie algebra Lie$\,{\cal G}$ and any representation space. 

Historically and according to the best of our knowledge, a preliminary construction was explicitly given in~\cite[see appendix F]{Laz90} or \cite{Sto93}. It could be viewed as a most formalized version of \cite[see pages 485 and 486]{Zum85} and exploits results in \cite[section II and appendix]{Sto85}. Latter, Raymond Stora guessed an homotopy after a thorough discussion at CERN-TH in the mid of the 1990's with one of us (SL) and evoked by Raymond Stora himself in a seminar given at CPT (Marseilles) in November 1995.

\smallskip
Let us give a short account on this construction mainly recorded in~\cite{Sto05} in order to bring this nice construction out of the shadows
as a part of the wide legacy left by Raymond Stora.

\medskip
To start with, let us denote the gauge group by ${\cal G} = \{g: U \rightarrow G\}$ the set of local maps with values in a compact, simple symmetry group $G$. It carries a group law inherited from that of $G$.
As field representation spaces for ${\cal G}$, one can distinguish 
\begin{itemize}
\item
when elements of $\cal G$ are considered as fields, one has the so-called gauge group ${\cal G} = \{\gamma \in {\cal G}, \gamma^g = g^{-1} \gamma g \}$, which is compatible with the group law in $G$. 

\item
the gauge transformation of gauge potentials with the usual action of $\cal G$ given by the right action
$a^\gamma = \gamma^{-1} a \gamma + \gamma^{-1} d\gamma.$
\end{itemize}
Consider a local consistency anomaly as an element in $H^1(\text{Lie}\,{\cal G},\Gamma_\text{\tiny loc}(a))$
\begin{align}
\label{WZconsist}
&A(c,a) = \int_M \Delta(c,a), &\text{fulfilling}\quad s A(c,a) = 0
\end{align}
where $a$ is a Yang-Mills like potential and $c$ the Faddeev-Popov ghost associated to the Lie algebra $\text{Lie}\,{\cal G}$ of the gauge group ${\cal G}$. It may be recalled that $\Delta(c,a)$ is a differential polynomial in $a$ ({\em i.e.} local in $a$ in the QFT sense), top form on $M$ which is linear in $c$.
The Wess-Zumino consistency condition $s A(c,a) = 0$ (local 1-cocycle) yields \cite{Sto84,DuboisViolette:1985jb}
\begin{align}
\label{WZloc}
s \Delta(c,a) + d \Delta'(c,a) =0
\end{align}
where $d$ is the de Rham differential on $M$.

For instance, in a Yang-Mills theory, the Slavnov operation $s$ is explicitly given by
\begin{align}
\label{BRS-alg}
& s a = - D_a c = -dc - [a,c], & &sc = -\tfrac{1}{2} [c,c], & s^2=0
\end{align}
where $[\, ,\,]$ is the graded Lie algebra bracket. Recall that $c\in (\text{Lie}\,\cal G)^*\otimes \text{Lie}\,\cal G$ and is the generator of cochains on $\text{Lie}\,\cal G$ \cite{Sto85} or \cite[section 6.10]{DeAzc-Izq}).

\medskip

\begin{thm}{\cite{WZ71}}
Given an anomaly $A(c,a)$ with $sA(c,a)=0$, one can construct a functional on the field space of gauge fields $a$ and on field $u\in \cal G$ 
\begin{align}
\label{WZ-Stora}
\Gamma_\text{\tiny WZ}(u,a) = \int_0^1 A(\gamma_t^{-1} d_t \gamma_t, a^{\gamma_t}) = \int_e^u A(\gamma^{-1} \delta \gamma, a^\gamma)
\end{align}
where $\{\gamma_t\}$ is a $1$-dimensional family in ${\cal G}$, for $t\in[0,1]$ and $\gamma_0=e$ and $\gamma_1=u$.\footnote{$e$ is the identity of ${\cal G}$ and denotes the constant field $x\mapsto e_G$.}
If $s$ is extended on $u$ by $s u = - c u$ such that $s(a^u)=0$ is secured, then 
\begin{align}
\label{sWZ}
s \Gamma_\text{\tiny WZ}(u,a) = A(c,a).
\end{align}
\end{thm} 

\medskip
The Wess-Zumino trick \cite{WZ71,Zumi84a} is to use a seesaw mechanism between left and right actions by extending the Slavnov operation $s$ on the gauge group $\cal G$ by
$s u = -cu$, where $u$ is considered as a field, in order to guarantee $s(a^u) = 0$. This is, in the BRST language, the infinitesimal version of the gauge invariance of a {\em composite field}
\[
a^u := \Ad(u^{-1}) a + u^{-1} d u
\] 
under the gauge transformations of both elementary fields $a$ and $u$ (the latter are considered as belonging to representation spaces of $\cal G$ --a right group action) according to
\begin{align}
\label{simul-gauge-trsf}
& a \rightarrow a^\gamma = \Ad(\gamma^{-1}) a + \gamma^{-1} d \gamma, & & 
u\rightarrow u^{\gamma} = \gamma^{-1} u 
& &\Rightarrow (a^u)^\gamma = (a^\gamma)^{u^\gamma} = (a^\gamma)^{\gamma^{-1} u} = a^u
\end{align}
where $\gamma$ is a gauge group element, $\gamma\in \cal G$. 
Let us stress that the composite field $a^u$ must not be considered as a gauge transformation of $a$ because $a^u$ is no longer a connection owing to the fact that $u$ carries a non-linear transformation law $u^{\gamma}= \gamma^{-1} u$ which is different from the required transformation law ${\gamma'}^\gamma = \gamma^{-1} \gamma' \gamma$ for a genuine gauge group element considered as a field in the `adjoint' representation, (as recalled above \eqref{WZconsist}). In fact, $a^u$ may be interpreted as a change of variable in the field space of the $(a,u)$'s; see discussion in \cite[see in particular section 2 in both of those references]{GaugeInv,FLM2015_II}
on what we called the {\em dressing field method}, a construction which goes back to Dirac \cite{Dirac55} and which, in turn, 
enters in the construction of the Wess-Zumino functional.

\paragraph{The Stora construction.} 
Given a family $\{\gamma_t\}$ in $\cal G$, $0\leq t\leq 1$, with $\gamma_0=e$ and $\gamma_1=u$, its action on $a$ gives a family $\{a_t :=a^{\gamma_t}\}$, interpolating from $a_0=a$ to $a_1=a^u$.
In the field space, consider the interpolating family $\{u_t,a_t\}$ defined by $(a_t)^{u_t} = a^u$ for all $t$. This constraint implies
$u_t = \gamma_t^{-1} u$ (in the case of a transitive action of $\cal G$), that is $u_t$ is a family of dressing fields with $u_0=u$ and $u_1=e$.

Requiring the invariance of $(a_t)^{u_t}$ under the gauge transformations given in \eqref{simul-gauge-trsf} infers
\[
(a_t)^{u_t} = (a^\gamma)^{\gamma^{-1} u} = ((a_t^{\gamma_t^{-1}})^\gamma)^{\gamma^{-1} \gamma_t u_t}
= (a_t^{\gamma_t^{-1} \gamma \gamma_t})^{\gamma_t^{-1}\gamma^{-1} \gamma_t u_t}.
\]
This shows the gauge invariance of $(a_t)^{u_t}$ under the following gauge transformations on the family 
\[
a_t \rightarrow a_t^{\gamma_t^{-1} \gamma \gamma_t}, \qquad 
u_t \rightarrow (\gamma_t^{-1} \gamma \gamma_t)^{-1} u_t.
\]
It is worthwhile to notice that $\gamma_t^{-1} \gamma \gamma_t$ is a family which stays within the gauge group $\cal G$ as a field space for the `adjoint' representation.
Let us turn to the infinitesimal version of the latter in a BRST language. To sum up, one has the family in field space
\[
u_t = \gamma_t^{-1} u, \qquad
a_t = a^{\gamma_t}, \qquad
c_t =  \gamma_t^{-1} c\, \gamma_t
\]
(the latter is the adjoint action of the family $\gamma_t$ on (Lie$\,{\cal G})^*$) and it can be checked that
\begin{align}
\label{BRS-family}
&s u_t = -c_t u_t, & & s a_t = -dc_t - c_t a_t - a_t c_t, & &
sc_t = - \tfrac{1}{2} [c_t,c_t], & & s^2=0.
\end{align}
Along some ideas given in \cite{ManeStorZumi85a}, the Stora's trick is to introduced a homotopy for $s$ on the family through an {\em even} derivation $k_t$ defined as
\begin{align}
\label{k-homotop}
k_t u_t = k_t a_t = 0, \qquad
k_t c_t = -d_t u_t u_t^{-1} = \gamma_t^{-1} d_t \gamma_t
\end{align}
in order to satisfies
\[
k_t s - s k_t = d_t, \qquad s^2 = d_t^2 = s d_t + d_t s = 0 
\]
where $d_t$ is an antiderivation along the 1-parameter family induced by $\gamma_t$.

Since the differential algebra \eqref{BRS-family} is similar to the BRST algebra \eqref{BRS-alg}, one has the consistency condition \eqref{WZconsist} along the whole family, namely, $s A(c_t,a_t) = 0$, or $s \Delta(c_t,a_t) + d \Delta'(c_t,a_t)= 0$. Therefore, by \eqref{k-homotop}, one gets
\begin{align*}
d_t A(c_t,a_t) = -s k_t A(c_t,a_t) = - s A(\gamma_t^{-1} d_t \gamma_t, a^{\gamma_t})
\end{align*}
where on the r.h.s. the integrand used in the Wess-Zumino-Stora formula \eqref{WZ-Stora} occurs.
Integration in $t$ yields 
\begin{align}
\label{int-consistency}
\int_0^1 d_t A(c_t,a_t) = A(c_1,a_1) - A(c_0,a_0) = A(u^{-1}c\, u, a^u) - A(c,a) =
- \int_0^1 s A(\gamma_t^{-1} d_t \gamma_t, a^{\gamma_t})
\end{align}
At this stage, some care is required, because $s$ acts on the upper integration bound $u$ also of the interpolating family $\{\gamma_t\}$. Hence, $s$ does not commute with the integration. The latter integration can be rewritten as
\[
- \int_0^1 s A(\gamma_t^{-1} d_t \gamma_t, a^{\gamma_t}) = - \int_e^u s A(\gamma^{-1}\delta \gamma, a^\gamma) = -s \int_e^u A(\gamma^{-1}\delta \gamma, a^\gamma) + s_{|u} \int_e^u A(\gamma^{-1}\delta \gamma, a^\gamma)
\]
where in the last term, only the upper bound is varied. Raymond Stora used to rewrite the latter as 
\[
s_{|u} \int_e^u A(\gamma^{-1}\delta \gamma, a^\gamma) = \left( \int_e^{u+su} - \int_e^u \right) A(\gamma^{-1}\delta \gamma, a^\gamma)
\]
in order to consider the difference between the integration of the anomaly along two different paths in $\cal G$, thanks to the possibility\footnote{This is at least possible in the connected component to the identity $e$ of $\cal G$, or if the topology of ${\cal G}$ is suitable for a vanishing fundamental group $\pi_1({\cal G})=0$, see \cite{Sto84}.} to smoothly deform $\{\gamma_t\}$ from $e$ to $u$ into a path from $e$ to $u+su=u-cu$ for a variation $\delta\gamma_t = - t c \gamma_t$.
Then, using \eqref{Stora-cor} of the Corollary in \ref{appendix:B} where a more detailed construction is given, one gets
\begin{align*}
s_{|u} \int_e^u A(\gamma^{-1}\delta \gamma, a^\gamma) = - A(\gamma^{-1}_1\delta \gamma_1, a^{\gamma_1}) = A(u^{-1}c u, a^u).
\end{align*}
Collecting all the various terms, formula \eqref{int-consistency} reduces to
\[
s \int_0^1 A(\gamma_t^{-1} d_t \gamma_t, a^{\gamma_t})  = A(c,a)
\]
a result which achieves the proof.

\section{More about deformation of interpolating families \label{appendix:B}}
\setcounter{equation}{0}

Let $\delta$ denote the differential on the gauge group ${\cal G}$ and the Maurer-Cartan form on $\cal G$ at $\gamma$ reads $\gamma^{-1} \delta\gamma$    \footnote{This field is the Faddev-Popov ghost according to the Zumino standpoint; see e.g. \cite{Bertlmann}.}. Following \cite{ManeStorZumi85a} and \cite[see appendix]{Sto85}, if 
\[
a^\gamma = \gamma^{-1} a \gamma + \gamma^{-1} d\gamma
\]
is the finite gauge transformed (obtained by right action) of the gauge field $a$ and since $\delta$ acts on the gauge group element $\gamma$ only, taking into account the Maurer-Cartan equation, one has the algebra
\begin{align}
\label{Zum-alg}
&\delta (a^\gamma) = - d (\gamma^{-1} \delta\gamma) - [a^\gamma,\gamma^{-1} \delta\gamma], & &\delta(\gamma^{-1} \delta\gamma) = -\tfrac{1}{2} [ \gamma^{-1} \delta\gamma\, ,\gamma^{-1} \delta\gamma],
& &\delta^2 = 0.
\end{align}
Note that this differential algebra \eqref{Zum-alg} is isomorphic to the BRS algebra given in~\eqref{BRS-alg}. This yields an homomorphism from $H^*(\text{Lie}\,{\cal G},\Gamma_\text{\tiny loc}(a))$ to $H^*_\delta ({\cal G},\Gamma_\text{\tiny loc}(a))$ \cite{Sto85,DuboisViolette:1985jb}. This shifts  cohomological issues on $\text{Lie}\,{\cal G}$ to the ones of left invariant forms on ${\cal G}$ with values in $\Gamma_\text{\tiny loc}(a)$. For instance, the replacements
\begin{equation}
\label{consistency-Zum}
\left. \begin{array}{l}
s \rightarrow \delta \\
c \rightarrow \gamma^{-1} \delta\gamma \\
a \rightarrow a^\gamma 
\end{array} \right\}
\Longrightarrow s A(c,a) = 0 \rightarrow \delta A(\gamma^{-1} \delta\gamma, a^\gamma) = 0,  \quad \text{ or } \delta \Delta(\gamma^{-1} \delta\gamma, a^\gamma) + d \Delta'(\gamma^{-1} \delta\gamma, a^\gamma)= 0.
\end{equation}
The formula on the r.h.s. shows that one is led to work with differential forms on $M\times \cal G$ \cite{DbV87proc}.
In particular, $A=\int_M \Delta$ is of degree 1 and is linear in $\gamma^{-1}\delta \gamma \in T^*\cal G \otimes \text{Lie}\,\cal G$.

\medskip
By exploiting further this correspondence, and thus the Wess-Zumino consistency condition on the 1-cocycle $A$, one can show the following important result. Consider an interpolating family $\{\gamma_t\}$, $0\leq t \leq 1$ in $\cal G$ from $\gamma_0 = e$ to $\gamma_1=g$. Restricting the 1-cocycle $A$ to this interpolating family, one gets by pulling-back the consistency condition \eqref{consistency-Zum} on $[0,1]$
\[
d_t A(\gamma_t^{-1} d_t \gamma_t,  a^{\gamma_t})=0.
\]
The Wess-Zumino action is a particular example of functionals over paths in $\cal G$ and given by
\[
\int_0^1  A(\gamma_t^{-1} d_t \gamma_t,  a^{\gamma_t}) = \int_{M\times[0,1]} 
(\tilde\gamma^{-1} \tilde\delta\tilde\gamma, a^{\tilde\gamma})
\]
by lifting the situation to paths $\tilde\gamma: M\times[0,1] \rightarrow G$, \cite[p.164]{ManeStorZumi85a}. 
In particular, it makes sense to integrate the 1-cochain $A=\int_M \Delta$ which is linear in $\gamma^{-1}\delta \gamma$.

\medskip
We address now the issue of the behavior of such functionals under smooth deformations of the interpolating family~$\{\gamma_t\}$ (as a 1-dimensional submanifold \footnote{Raymond Stora had in mind that it was possible to extend the deformation to any submanifold in $\cal G$.} of $\cal G$). Deforming the family $\{\gamma_t\}$ amounts to defining a map 
\begin{align}
\label{deform}
\hat\gamma: [0,1] \,\times\, ]-\epsilon,\epsilon\,[ \rightarrow {\cal G}, \quad
(t,\tau) \mapsto \hat\gamma (t,\tau) = \gamma_{t,\tau} 
\quad \text{with } \gamma_{t,\tau=0} = \gamma_t.
\end{align}
The smooth deformation $\tau \mapsto \gamma_{t,\tau}$ defines, for each $t$, a curve in $\cal G$ passing through $\gamma_t$ at $\tau=0$ with velocity $\frac{\partial}{\partial \tau} \left( \gamma_{t,\tau}\right)_{|\tau=0}$. This gives rise to a $t$-dependent family of tangent vectors (as it will be explicitly seen later on)
defines along the family $\{\gamma_t\}$ by
\[
X_{|\gamma_t} = \frac{\partial}{\partial \tau} \left( \gamma_{t,\tau}\right)_{|\tau=0} \in  T_{\gamma_t} {\cal G}
\]
where $T_{\gamma_t} {\cal G}$ is the tangent space to $\cal G$ at the point $\gamma_t$. 
We are interested in computing the variation of the smooth map
\[
\tau \mapsto \int_0^1  A(\gamma_{t,\tau}^{-1} d_t \gamma_{t,\tau},  a^{\gamma_{t,\tau}})
\]
namely, one has indeed to compute one of the derivatives
\begin{align}
\label{der-int}
\text{either} & & \frac{\partial}{\partial \tau} \left( \int_0^1  A(\gamma_{t,\tau}^{-1} d_t \gamma_{t,\tau},  a^{\gamma_{t,\tau}}) \right)_{|\tau=0} & & \text{or} & & \frac{\partial}{\partial \tau} \left( \int_0^1  \Delta(\gamma_{t,\tau}^{-1} d_t \gamma_{t,\tau},  a^{\gamma_{t,\tau}}) \right)_{|\tau=0}.
\end{align}
The derivative on the r.h.s. allows to work easier with differential forms on $M\times \cal G$.
In this respect, interesting developments might be found in \cite[p.192ff.]{Iglesias2013}. By smoothness, one has
\[
\frac{\partial}{\partial \tau} \left( \int_0^1  \Delta(\gamma_{t,\tau}^{-1} d_t \gamma_{t,\tau},  a^{\gamma_{t,\tau}}) \right)_{|\tau=0} = \int_0^1 \frac{\partial}{\partial \tau} \left( \Delta(\gamma_{t,\tau}^{-1} d_t \gamma_{t,\tau},  a^{\gamma_{t,\tau}}) \right)_{|\tau=0}.
\]
It is useful to work at the level of generators of the differential algebra, see e.g. \cite{DbV87proc}. One can check, by virtue of \eqref{Zum-alg} and by $(\delta \gamma_t)(X_{|\gamma_t}) = X_{|\gamma_t}$, that
\begin{align}
\label{var-gen}
\frac{\partial}{\partial \tau} \left( a^{\gamma_{t,\tau}} \right)_{|\tau=0} &= \frac{\partial}{\partial \tau} \left( 
\gamma_{t,\tau}^{-1} a\, \gamma_{t,\tau} + \gamma_{t,\tau}^{-1} d \gamma_{t,\tau} \right)_{|\tau=0} = \left(\delta (a^{\gamma_t}) \right) (X_{|\gamma_t}), \notag\\[-2mm]
& \\[-2mm]
\frac{\partial}{\partial \tau} \left(\gamma_{t,\tau}^{-1} d_t \gamma_{t,\tau} \right)_{|\tau=0} &= d_t \left( \gamma_t^{-1} (\delta \gamma_t) (X_{|\gamma_t}) \right) + [\gamma_t^{-1} d_t \gamma_t ,\gamma_t^{-1} (\delta \gamma_t) (X_{|\gamma_t}) ]. \notag
\end{align}
Dropping out the tangent vectors $X_{|\gamma_t}$ yields that the l.h.s. derivative in \eqref{der-int} comes down to computing
\begin{align}
\label{del-int}
\delta \int_0^1  A(\gamma_t^{-1} d_t \gamma_t,  a^{\gamma_t}) = \int_0^1  \delta A(\gamma_t^{-1} d_t \gamma_t,  a^{\gamma_t})
\end{align}
and \eqref{var-gen} leads to
\begin{align*}
\delta \left( a ^{\gamma_t}\right) = -d \left(\gamma_t^{-1} \delta \gamma_t \right) - [ a ^{\gamma_t} ,\gamma_t^{-1} \delta \gamma_t ], \qquad
\delta \left( \gamma_t^{-1} d_t \gamma_t \right) = - d_t \left(\gamma_t^{-1} \delta \gamma_t \right) - [ \gamma_t^{-1} \delta \gamma_t, \gamma_t^{-1} d_t \gamma_t ]
\end{align*}
for any variation $\{\delta \gamma_t\}$  of the family. Moreover, one has also (morally, it corresponds to the pull-back on $[0,1] \,\times\, ]-\epsilon,\epsilon\,[\,$, see \cite{Iglesias2013})
\begin{align*}
(d_t + \delta) \left( a ^{\gamma_t}\right) &= -d \left(\gamma_t^{-1} (d_t + \delta) \gamma_t \right) - [ a ^{\gamma_t} ,\gamma_t^{-1} (d_t + \delta) \gamma_t ] 
\end{align*}
together with the Maurer-Cartan equation
\[
(d_t + \delta) \left( \gamma_t^{-1} (d_t + \delta) \gamma_t \right) 
+ \tfrac{1}{2} [ \gamma_t^{-1} (d_t + \delta) \gamma_t, \gamma_t^{-1} (d_t + \delta) \gamma_t ] =0.
\]
The last two equations reproduce a differential algebra similar to 
the BRS algebra~\eqref{BRS-alg} and accordingly the cocycle condition \eqref{WZconsist} on $A$ yields
\begin{align}
\label{WZ-I}
(d_t + \delta) A(\gamma_t^{-1} (d_t + \delta) \gamma_t,  a^{\gamma_t}) = 0.
\end{align}
By polarization, one gets the important condition
\begin{align*}
\delta A(\gamma_t^{-1} d_t \gamma_t,  a^{\gamma_t}) + d_t  A(\gamma_t^{-1} \delta \gamma_t,  a^{\gamma_t}) = 0.
\end{align*}
which once inserted into \eqref{del-int} gives 
\footnote{Adapting to our context the approach given in \cite[§6.70, 6.71]{Iglesias2013} corresponds to 
\[
\delta \int_{[0,1]} A(\gamma_t^{-1} d_t \gamma_t,  a^{\gamma_t}) = \int_{[0,1]} (d_t+ \delta) A(\gamma_t^{-1} (d_t + \delta) \gamma_t,  a^{\gamma_t}) - \int_{[0,1]} d_t  A(\gamma_t^{-1} \delta \gamma_t,  a^{\gamma_t}) = - \int_{\partial[0,1]} A(\gamma_t^{-1} \delta \gamma_t,  a^{\gamma_t})
\]
and due to \eqref{WZ-I}, only the integral on the boundary contributes in the r.h.s.}
\begin{align}
\label{bornes}
\delta \int_0^1  A(\gamma_t^{-1} d_t \gamma_t,  a^{\gamma_t}) =  - \int_0^1 d_t  A(\gamma_t^{-1} \delta \gamma_t,  a^{\gamma_t}) = 
A(\gamma_0^{-1} \delta \gamma_0,  a^{\gamma_0}) - A(\gamma_1^{-1} \delta \gamma_1,  a^{\gamma_1}).
\end{align}
This shows that the variation depends on the integration limits only. One has thus proved the

\begin{lem} [Stora]
An infinitesimal deformation of the interpolation depends only of the variation at the ends of the path:
\begin{align}
\label{Stora-lem}
\delta \int_0^1 A(\gamma_t^{-1} d_t \gamma_t,  a^{\gamma_t}) = A(\gamma_0^{-1} \delta \gamma_0,  a^{\gamma_0}) - A(\gamma_1^{-1} \delta \gamma_1,  a^{\gamma_1}).
\end{align}
\end{lem}
In particular, if $\delta \gamma_0 = \delta \gamma_1=0$, this result shows the independence on the choice of the interpolating family $\gamma_t$ between $e$ and $g$ for computing $\int_0^1 A(\gamma_t^{-1} d_t \gamma_t, a^{\gamma_t})$.
If the topology of ${\cal G}$ is suitable, e.g. the fundamental group $\pi_1({\cal G})=0$, the consistency condition \eqref{consistency-Zum} can be recast as a coboundary condition on ${\cal G}$, and by Stokes theorem
\[
\int_{S} \delta A(\gamma^{-1} \delta\gamma,  a^\gamma) = 0 = \oint_{\text{loop}} A(\gamma^{-1} \delta\gamma,  a^\gamma)
\]
for $S$ a surface in ${\cal G}$ enclosed by a loop passing through $e$ and $g$.
\smallskip
This result is to some extent the infinitesimal version of the group cohomology introduced in~\cite{Zum85}.

\medskip
%

The deformation of the family $\{\gamma_t\}$ considered in 
\ref{WZS} with one end kept fixed, is achieved by choosing for \eqref{deform} the left action, $\gamma_{t,\tau} = e^{-\tau t \chi} \gamma_t$, (as a smooth homotopy)
with $\chi \in \text{Lie}\,{\cal G}\simeq T_e {\cal G}$. 
One readily checks that this choice for $\gamma_{t,\tau}$  gives, on the one hand, $\gamma_{0,\tau} = \gamma_0=e$ for $t=0$ and for any $\tau$, and, on the other hand, $\gamma_{1,\tau} = e^{- \tau X} \gamma_1$ for $t=1$. With such a smooth deformation, the induced vector field along the family $\{\gamma_t\} \subset \cal G$ is given by
\[
X_{|\gamma_t} = \frac{\partial}{\partial \tau} \left( \gamma_{t,\tau}\right)_{|\tau=0} = - t\,T_e R_{\gamma_t} \chi
\]
where $T_e R_\gamma$ is the tangent map of the right translation. The eveluation of the Maurer-Cartan form gives
\[
(\gamma_t^{-1}\delta \gamma_t)(X_{|\gamma_t}) = - t\,T_{\gamma_t} L_{\gamma_t^{-1}} T_e R_{\gamma_t} \chi = - t\, T_e (L_{\gamma_t^{-1}} \circ R_{\gamma_t}) \chi = -t\, \gamma_t^{-1} \chi \gamma_t
\]
which reflects (up to a sign) the right-equivariance of the Maurer-Cartan form on $\cal G$:
$(\gamma^{-1}\delta \gamma)(T_e R_\gamma \chi) = \, \gamma^{-1} \chi \gamma$ for $\chi\in\text{Lie}\,\cal G$.
One has $\delta \gamma_0 = 0$ and $(\delta \gamma_1) (X_{|\gamma_1}) = - \chi \gamma_1$.
Using the algebraic definition of the Faddeev-Popov ghost \cite{Sto85}, as the ``$\text{Lie}\,\cal G$-valued generator of $(\text{Lie}\,{\cal G})^*$'', namely, $c(\chi)=\chi$, one can write $\delta \gamma_1  = - c\,\gamma_1$. Combining this construction with the above Lemma, and with a slight abuse of notation, one has the 
 
\begin{cor} [Stora]
For an infinitesimal deformation $\{\delta \gamma_t = - t c\, \gamma_t\}$ of the interpolating family $\{\gamma_t\}$ from $\gamma_0 = e$ to $\gamma_1$, one has the variation
\begin{align}
\label{Stora-cor}
\delta \int_0^1 A(\gamma_t^{-1} d_t \gamma_t,  a^{\gamma_t}) = A(\gamma_1^{-1} c\, \gamma_1,  a^{\gamma_1}).
\end{align}
\end{cor}


\end{document}